\documentclass[11pt,a4paper]{article}

\usepackage{amsmath,amssymb,cite,epsfig,mathrsfs}

\def\scri{\mathscr{J}}

\begin{document}

\title{Pure double-layer bubbles in quadratic $F(R)$ gravity} 
\author{Ernesto F. Eiroa$^{1, 2}$\thanks{e-mail: eiroa@iafe.uba.ar}, Griselda Figueroa Aguirre$^{1}$\thanks{e-mail: gfigueroa@iafe.uba.ar}, and Jos\'e M. M. Senovilla$^{3}$\thanks{e-mail: josemm.senovilla@ehu.eus}\\
{\small $^1$ Instituto de Astronom\'{\i}a y F\'{\i}sica del Espacio (IAFE, CONICET-UBA),}\\
{\small Casilla de Correo 67, Sucursal 28, 1428, Buenos Aires, Argentina}\\
{\small $^2$ Departamento de F\'{\i}sica, Facultad de Ciencias Exactas y 
Naturales,} \\ 
{\small Universidad de Buenos Aires, Ciudad Universitaria Pabell\'on I, 1428, 
Buenos Aires, Argentina}\\
{\small $^3$ F\'isica Te\'orica, Universidad del Pa\'is Vasco, Apartado 644, 48080 Bilbao, Spain.}} 

\maketitle

\begin{abstract}
We present a class of spherically symmetric spacetimes corresponding to bubbles separating two regions with constant values of the scalar curvature, or equivalently with two different cosmological constants, in quadratic $F(R)$ theory. The bubbles are obtained by means of the junction formalism, and the matching hypersurface supports in general a thin shell and a gravitational double layer. In particular, we find that pure double layers are possible for appropriate values of the parameters of the model whenever the quadratic coefficient is negative. This is the first example of a pure double layer in a gravitational theory.
\end{abstract}

\section{Introduction}

Gravitational fields are created by masses (and energies), and only positive masses/energies have been observed in nature. Therefore, the existence of mass dipoles, or of shells with mass dipole distributions, was not expected in gravitational theories. It thus came as a surprise the demonstration that singular distributions describing the analogues of electrostatic dipole layers are actually possible in gravitational theories with a Lagrangian quadratic in the curvature \cite{js1,js2,js3,rsv}. This was first proven in the so-called $F(R)$ theories with a careful analysis of  the proper junction conditions \cite{dss,js1}, and after noticing \cite{js1} that there is an exceptional possibility ---when the function $F(R)$ is quadratic--- allowing for the existence of shells supporting a ``Dirac-delta-prime" type of distribution. A more complete analysis along with the first examples of these {\em gravitational double layers} was given in \cite{js2,js3}. The junction conditions were extended in the past year to the most general gravitational theory with a Lagrangian quadratic in the curvature \cite{rsv}, proving in particular that the possible presence of gravitational double layers is a shared feature for all of them --with some peculiarities for the $F(R)$ case.

In $F(R)$ theories \cite{revfr1,revfr2,revfr3}, the scalar curvature $R$ corresponding to the Einstein-Hilbert Lagrangian is replaced in the gravitational action by a function $F(R)$. This kind of models can provide a unified picture of both inflation in the early universe and the accelerated expansion observed at later times. Within $F(R)$ gravity, spherically symmetric black hole solutions have been found in recent years \cite{bhfr1,bhfr2,bhfr3}. Lorentzian wormhole geometries \cite{whfr1,whfr2,whfr3,whfr4}, collapsing spherical stars \cite{gnmg}, and thin shell wormholes \cite{gfa1,gfa2} have all been studied in these theories. The particular case of quadratic $F(R)$ gravity, which has a positive mass theorem \cite{stro} and a well-defined entropy formulation \cite{jkm}, can provide a self-consistent model for inflation \cite{sta}. Gravitational radiation \cite{nj} and realistic compact relativistic (neutron and quark) stars \cite{acdo} have also been investigated for Lagrangians quadratic in the scalar curvature. 

From the junction condition point of view, a non-linear $F(R)$ always requires the equality of the trace of the second fundamental forms at both sides of the joining hypersurface and, with the exception of quadratic $F(R)$, the continuity of the scalar curvature $R$. However, in quadratic $F(R)$, the scalar curvature can be discontinuous at the matching hypersurface, resulting in a much richer structure of the matter contents supported on it: besides the standard energy-momentum tensor, an external energy flux vector, an external scalar tension (or pressure), and another energy-momentum contribution ---resembling dipole distributions in classical electrodynamics--- arise \cite{js1,js2}. All these contributions are required to make the whole energy-momentum tensor divergence free \cite{js1,js2}. This dipole distribution can be interpreted as a gravitational double layer. 

A very interesting and radical possibility was proposed in \cite{js2,js3}: the existence of {\em pure} double layers, characterized by having, at the matching hypersurface, vanishing hypersurface energy-momentum tensor, external energy flux vector and external scalar tension, but a non-vanishing dipole strength. However, no explicit examples had been found up to now. Herein, in quadratic $F(R)=R-2\Lambda+\alpha R^2$, we present a class of spacetimes with spherical symmetry which can have, for suitable values of the parameters, a pure double layer matching the inner and the outer regions of a four dimensional manifold. This is the first example of such an object, and the outer geometry is exactly that of a Kottler ---or Schwarzschild-de Sitter--- spacetime with positive cosmological constant. The inner region, on the other hand, is just a portion of de Sitter spacetime (with a different cosmological constant). The mass parameter of the outer spacetime is directly linked to the strength of the pure double layer, which in turn is related to the deficit of the external cosmological constant ---with respect to the inner one. A thorough analysis of the properties of the spacetime is presented, along with a discussion of the interpretation of the results.

The plan is as follows: in Sec. \ref{setting} we provide the basic junction conditions in quadratic $F(R)$ and the field equations of the hypersurface energy-momentum quantities. Then in Sec. \ref{bubbles}, we present the general construction of static thin shells with a double layer, which can be understood as bubbles with constant area. In Sec. \ref{pure}, we show the conditions for which a pure double layer is found. We prove that, for a fixed negative value of $\alpha$, the cosmological constant on one side of the pure double layer can be chosen between some definite limits, and then the cosmological constant on the other side, the area of the double layer, and the mass parameter of the outer spacetime are all uniquely determined. We also provide some conformal diagrams of the pure double layer spacetimes. Finally, in Sec. \ref{discu}, we summarize and discuss the results obtained. In this article, we adopt units so that $G=c=1$.

\section{Setting: junction conditions in quadratic $F(R)$}\label{setting}

In a junction problem (see e.g. \cite{ms}), the manifold $\mathcal M$ has an inner part $\mathcal M_1$ and an outer part $\mathcal M_2$, with the gluing hypersurface $\Sigma$ corresponding to a boundary hypersurface or a thin shell, depending on the junction conditions. The first fundamental form on $\Sigma$ is denoted by $h_{\mu \nu}$, and the second fundamental forms (or extrinsic curvatures) by $K^{1,2}_{\mu \nu}$, where the superscripts ---here and elsewhere, and occasionally subscripts too--- refer to the inner and outer parts of $\mathcal M$. The jump of any quantity $\Upsilon $ across $\Sigma$ is denoted by 
$$[\Upsilon ]\equiv (\Upsilon ^{2}-\Upsilon  ^{1})|_\Sigma .
$$
(Observe that interchanging $1 \longleftrightarrow 2$ changes the sign of these jumps). We have not added superscripts to the first fundamental form because, in order to have a well-defined curvature ---in the distributional sense--- even at $\Sigma$, one has to demand (as in general relativity) the agreement of the first fundamental forms inherited from both $\mathcal M_{1,2}$, that is,  $[h_{\mu \nu }]=0$ \cite{ms}. The unit normal $n^\mu$ to $\Sigma$ is, as a consequence, well defined without a jump at $\Sigma$, despite the fact that for computational purposes one usually needs to give its two expressions at both sides of $\Sigma$.

In generic $F(R)$ theories there are additional necessary conditions \cite{js1}:  the trace of the second fundamental form cannot have a jump
$$[K^{\mu}_{\;\; \mu}]=0
$$
and, whenever  $d^3F(R)/dR^3 \neq 0$, the continuity of $R$ across the joining surface is required, i.e. $[R]=0$ \cite{js1}. However, in any quadratic theory 
\begin{equation}
F(R)=R -2\Lambda +\alpha R^2 \label{quadF}
\end{equation}
where $\alpha$ is the parameter selecting the particular theory, a discontinuity of $R$ at the hypersurface $\Sigma$ is permitted \cite{js1,js2,js3}. The energy-momentum quantities supported at $\Sigma$ in this quadratic case take the explicit expressions \cite{js1,rsv}
\begin{equation}
\kappa S_{\mu \nu} =-[K_{\mu\nu}]+2\alpha( [n^{\gamma }\nabla_{\gamma}R] h_{\mu\nu}-[RK_{\mu\nu}]),  \qquad  n^{\mu}S_{\mu\nu}=0,
\label{LanczosQuad}
\end{equation}
where $\kappa =8\pi $, $S_{\mu \nu}$ represents the standard hypersurface energy-momentum tensor, and $\nabla$ is the covariant derivative; besides this there are also three other contributions: an external energy flux vector
\begin{equation}
\kappa\mathcal{T}_\mu=-2\alpha \overline{\nabla}_\mu[R],  \qquad  n^{\mu}\mathcal{T}_\mu=0,
\label{Tmu}
\end{equation}
where $\overline{\nabla }$ is the intrinsic covariant derivative on $(\Sigma,h_{\mu\nu})$; an external scalar pressure or tension
\begin{equation}
\kappa\mathcal{T}=2\alpha [R] K^\gamma{}_\gamma ;
\label{Tg}
\end{equation}
and a two-covariant symmetric tensor distribution 
\begin{equation}
\kappa \mathcal{T}_{\mu \nu}=\nabla_{\gamma } \left( 2\alpha [R] h_{\mu \nu } n^{\gamma } \delta ^{\Sigma }\right),
\label{dualay1}
\end{equation}
where $\delta ^{\Sigma }$ is the Dirac delta with support on $\Sigma $, or equivalently\footnote{Observe that there is an error in this formula in \cite{js1,js2,js3}, the indices in $\Psi^{\mu\nu}$ were missing.}
\begin{equation}
\kappa \left<\mathcal{T}_{\mu \nu},\Psi ^{\mu \nu } \right> = -\int_\Sigma 2\alpha[R] h_{\mu \nu }  n^\gamma\nabla_\gamma \Psi ^{\mu \nu },
\label{dualay2}
\end{equation}
for any test tensor field $\Psi ^{\mu \nu }$. This {\em double layer} energy-momentum distribution $\mathcal{T}_{\mu \nu }$ corresponds to a Dirac ``delta prime'' type contribution with strength \cite{js1,rsv}
\begin{equation}
\kappa \mathcal{P}_{\mu \nu } =2\alpha[R] h_{\mu \nu }, \hspace{1cm}  \mathcal{P}_{\mu \nu } = \mathcal{P}_{\nu \mu }, \hspace{1cm} n^{\mu } \mathcal{P}_{\nu \mu } =0 \label{strength}
\end{equation}
resembling dipole distributions in classical electrodynamics \cite{js1,js2,js3,rsv} .

All the above contributions are required in order to make the complete energy-momentum tensor divergence free \cite{js1,js2,rsv}, which is necessary for local conservation. If $K_{\mu \nu}$ and $R$ have no jumps at $\Sigma$, all these contributions vanish and $\Sigma $ is an ordinary boundary hypersurface. In general, there is a thin shell plus a double layer at the matching hypersurface. When $S_{\mu \nu} \neq 0$ and $[R] =0$, we have only the usual thin shell, but if $S_{\mu \nu} = 0$, $\mathcal{T}_\mu =0$, and $\mathcal{T}=0$ but $[R] \neq 0$, we obtain a pure double layer.

The previous quantities also satisfy the following identities and field equations \cite{js2,js3,rsv}:
\begin{equation}
\kappa \left( S^\rho{}_\rho +\mathcal{T}\right)=6\alpha [n^{\gamma }\nabla_{\gamma}R] ,\label{trace}
\end{equation}
\begin{equation}
\mathcal{T}_\mu =-\overline\nabla^\rho \mathcal{P}_{\rho\mu}, \label{divP}
\end{equation}
\begin{equation}
\mathcal{T} = K_\Sigma^{\mu\nu} \mathcal{P}_{\mu\nu} =K^{\mu\nu} \mathcal{P}_{\mu\nu} ,\label{KP}
\end{equation}
\begin{equation}
n^\rho h^\sigma_\mu [T_{\rho\sigma}] +\overline\nabla^\rho S_{\rho\mu}
=-K^\rho{}_\rho \mathcal{T}_\mu -\overline\nabla_\mu \mathcal{T} ,\label{divS}
\end{equation}
\begin{equation}
\kappa\left( n^\rho n^\sigma [T_{\rho\sigma}] -K_\Sigma^{\rho\sigma}S_{\rho\sigma} +\overline\nabla^\rho\mathcal{T}_\rho \right) = 2\alpha [R] \left(R^\Sigma_{\rho\sigma}n^\rho n^\sigma +K^\Sigma_{\rho\sigma} K_\Sigma ^{\rho\sigma}  \right) 
\label{SK}
\end{equation}
where $T_{\mu\nu}$ is the energy-momentum tensor of the spacetime and the superscript $\Sigma$ means that the quantity must be evaluated at the matching hypersurface, and for ``discontinuous" quantities this means, for instance, for the Ricci tensor
$$
R^\Sigma_{\rho\sigma} =\frac{1}{2} \left(R^1_{\rho\sigma} +R^2_{\rho\sigma} \right)|_\Sigma .
$$
From these relations one immediately deduces that {\em pure} double layers require the following simultaneous conditions, keeping $[R]\neq 0$:
\begin{equation}
K^\rho{}_\rho =0, \hspace{3mm} \overline\nabla_\mu [R] =0,  \hspace{3mm}  [n^\gamma \nabla_\gamma R]=0,  \hspace{3mm} n^\rho h^\sigma_\mu [T_{\rho\sigma}]=0, \label{pure1}
\end{equation}
\begin{equation}
n^\rho n^\sigma [T_{\rho\sigma}]  = 2\frac{\alpha}{\kappa} [R] \left(R^\Sigma_{\rho\sigma}n^\rho n^\sigma +K^\Sigma_{\rho\sigma} K_\Sigma ^{\rho\sigma}  \right), \hspace{3mm} [K_{\mu\nu}]+2\alpha[RK_{\mu\nu}]=0.\label{pure2}
\end{equation}
The first equation in (\ref{pure1}) implies that $\Sigma$ should have zero mean curvature, while the second and third ones are automatically satisfied if the scalar curvatures $R_{1,2}$ are constants on $\Sigma$. The last in (\ref{pure1}) requires that the tangent-normal components of the energy-momentum tensor are continuous on $\Sigma$, which means no discontinuity on the fluxes of heat/energies. Of course, this is trivially satisfied if we are dealing with vacuum solutions. As observed in \cite{js2,js3} it follows that, if we choose a minimal hypersurface as matching $\Sigma$ in a spacetime with constant scalar curvatures on both $\mathcal{M}_{1,2}$ and vanishing energy-momentum tensors $T^{1,2}_{\mu\nu}$, then all conditions (\ref{pure1}) are automatically satisfied and we must only care about the two relations (\ref{pure2}) ---appropriately restricted. The meaning of these two equations is that of energy-momentum balance including the pure double layer, due to the very specific form of its strength (\ref{strength}) proportional to the first fundamental form (for details, see \cite{rsv}).

\section{Spherical bubbles}\label{bubbles}

We first present a general class of spherically symmetric geometries in quadratic $F(R)$, constructed by matching two manifolds with constant scalar curvatures, and we then provide a specific example representing a bubble.

\subsection{General construction}\label{gencons}

We begin our construction with two static and spherically symmetric spacetimes of the form, in standard coordinates, 
\begin{equation}
ds_{1,2}^2 = -A_{1,2}(r_{1,2})dt_{1,2}^2 +A_{1,2}^{-1}(r_{1,2})dr_{1,2}^2+r_{1,2}^2(d\theta^2+\sin^2\theta d\varphi^2),
\label{metric}
\end{equation}
where $A_{1,2}(r_{1,2})$ are functions of only the area coordinates $r_{1,2}$ on each side, 
with $r_{1,2}\ge 0$, $0 \le \theta < \pi$, and $0 \le \varphi < 2 \pi$. In both spacetimes we choose corresponding spherically symmetric hypersurfaces $\Sigma^{1,2}$ with fixed constant area coordinates $r_{1,2} =a_{1,2}$, and then we select two manifolds $\mathcal{M}_1$ and $\mathcal{M}_2$, defined by $0\le r_1 \leq a_1$ and $r_2\geq a_2$, respectively. We mathematically construct a new manifold $\mathcal M$ as the union of $\mathcal{M}_1$ and $\mathcal{M}_2$ with the obvious identification of points at $\Sigma^1$ and $\Sigma^2$ . By identifying in a natural way the angular coordinates everywhere, the coordinates of the embedding are $X_{1,2}^\mu=(t_{1,2},r_{1,2},\theta,\varphi)$ while the intrinsic coordinates ---now unique after identification--- at the hypersurface $\Sigma$ are $\xi ^i=(\tau , \theta,\varphi )$, with $\tau$ proper time on $\Sigma$. The induced metrics from both sides at the joining hypersurface $\Sigma$ read
$$
d\gamma_{1,2}^2=-A_{1,2}(a) \left(\frac{dt_{1,2}}{d\tau}\right)^2 d\tau^2+a_{1,2}^2 (d\theta ^2+\sin^2\theta d\varphi ^2)=-d\tau^2+a_{1,2}^2 (d\theta ^2+\sin^2\theta d\varphi ^2),
$$
so that the equality of the first fundamental forms then implies, on the one hand
$$
a_1 = a_2 := a
$$
and, on the other hand (fixing the free signs by choosing all times $t_{1,2}$ and $\tau$ to run to the future),
$$
\sqrt{A_1(a)}\,  \frac{dt_1}{d\tau} = \sqrt{A_2(a)} \, \frac{dt_2}{d\tau} .
$$
We adopt at the shell the orthonormal basis $\{ e_{\hat{\tau}}=e_{\tau }, e_{\hat{\theta}}=a^{-1}e_{\theta }, e_{\hat{\varphi}}=(a\sin \theta )^{-1} e_{\varphi }\} $. In this frame, the first fundamental form is simply $h_{\hat{\imath}\hat{\jmath}}= \mathrm{diag}(-1,1,1)$. The unit normal ($n^{\gamma }n_{\gamma }=1$) pointing from $\mathcal{M}_1$ to $\mathcal{M}_2$ has the expression, on each side of $\Sigma$, 
\begin{equation}
n_{\gamma }^{1,2 }= \left(0,\frac{1}{\sqrt{A_{1,2}(a)}},0,0 \right) \label{normal} .
\end{equation}

The second fundamental forms on both sides can be computed by using
\begin{equation}
K_{ij}^{1,2 }=-n_{\gamma }^{1,2 }\left. \left( \frac{\partial ^{2}X^{\gamma
}_{1,2} } {\partial \xi ^{i}\partial \xi ^{j}}+\stackrel{1,2}{\Gamma}{}_{\alpha \beta }^{\gamma }
\frac{ \partial X^{\alpha }_{1,2}}{\partial \xi ^{i}}\frac{\partial X^{\beta }_{1,2}}{
\partial \xi ^{j}}\right) \right| _{\Sigma },
\label{sff}
\end{equation}
so that we obtain for their non-vanishing components
\begin{equation} 
K_{\hat{\theta}\hat{\theta}}^{1,2 }=K_{\hat{\varphi}\hat{\varphi}}^{1,2
}= \frac{\sqrt{A_{1,2} (a) }}{a},
\label{Kth}
\end{equation}
and
\begin{equation} 
K_{\hat{\tau}\hat{\tau}}^{1,2 }= - \frac{A'_{1,2}(a)}{2 \sqrt{A_{1,2}(a)}},
\label{Ktau}
\end{equation}
with the prime representing the derivative with respect to their corresponding arguments (i.e. $r_1$ or $r_2$).  By using Eqs. (\ref{Kth}) and (\ref{Ktau}), the condition $[K^{\hat{\imath}}_{\;\; \hat{\imath}}]=0$ takes the form
\begin{equation} 
\frac{a A_{1}'(a) + 4A_{1}(a)}{\sqrt{A_{1}(a)}}=\frac{a A_{2}'(a)+ 4A_{2}(a)}{\sqrt{A_{2}(a)}}.
\label{CondGen}
\end{equation}

From now on, and for the purpose of the explicit construction carried out in the next subsection, we assume that the scalar curvatures $R_1$ and $R_2$ are both constant, but can be different from one another. In this case, Eq. (\ref{LanczosQuad}) simplifies to
\begin{equation}
\kappa S_{\hat{\imath}\hat{\jmath} } =-[K_{\hat{\imath}\hat{\jmath} }]-2\alpha[RK_{\hat{\imath}\hat{\jmath} }].
\label{LanczosGen2}
\end{equation}
In the chosen orthonormal frame, this Eq. (\ref{LanczosGen2}) implies that 
$S_{\hat{\imath}\hat{\jmath} }={\rm diag}(\sigma ,p,p)$
where $\sigma$ is the hypersurface energy density given by  
\begin{equation} 
\sigma = -\frac{A'_{1}(a)}{2\kappa\sqrt{A_{1}(a)}}\left(1+2\alpha R_{1}\right)+\frac{A'_{2}(a)}{2\kappa\sqrt{A_{2}(a)}}\left(1+2\alpha R_{2}\right) ,
\label{enden}
\end{equation}
and $p$ is the isotropic pressure given by
\begin{equation}
p = \frac{\sqrt{A_{1}(a)}}{\kappa a} \left(1+2\alpha R_{1} \right)-\frac{\sqrt{A_{2}(a)}}{\kappa a} \left(1+2\alpha R_{2} \right).
\label{pres}
\end{equation}
From Eq. (\ref{Tmu}) we easily obtain that $\mathcal{T}_\mu =0$, i.e. a vanishing external energy flux vector. The external scalar tension/pressure $\mathcal{T}$ has the form
\begin{equation}
\mathcal{T} = \frac{a A'_{1}(a) + 4 A_{1}(a)} {\kappa a \sqrt{A_{1}(a)}} \alpha [R] =  \frac{a A'_{2}(a) + 4 A_{2}(a)} {\kappa a \sqrt{A_{2}(a)}} \alpha [R],
\label{T}
\end{equation}
which by using Eq. (\ref{CondGen}) can be rewritten as
\begin{equation} 
\mathcal{T} = -\frac{ a A_{1}'+4 A_{1}(a)}{\kappa a\sqrt{A_{1}(a) }}\alpha R_{1}+\frac{a A_{2}'+4 A_{2}(a)}{\kappa a\sqrt{A_{2}(a) }}\alpha R_{2}.
\label{Trewrit}
\end{equation}
From Eqs. (\ref{enden}), (\ref{pres}), and (\ref{Trewrit}), it is easy to see that $\sigma$, $p$, and $\mathcal{T}$ are related by the equation of state $\sigma-2p=\mathcal{T}$, which is nothing but the expression in our case of Eq.  (\ref{trace}). We also have the dipole distribution, with strength
\begin{equation}
\kappa \mathcal{P}_{\hat{\imath}\hat{\jmath}} = 2\alpha[R] h_{\hat{\imath}\hat{\jmath}},
\label{dipo}
\end{equation}
which is non zero when $R_1 \neq R_2$.

\subsection{An explicit spherical bubble}

To provide a concrete example of a bubble, we adopt a well known geometry for our construction. The field equations, in the metric formalism, corresponding to Eq.  (\ref{quadF}) and with a constant scalar curvature $R$ admit the spherically symmetric solution of the form given by Eq. (\ref{metric}), in which the metric function reads  \cite{bhfr2} 
\begin{equation} 
A(r) = 1-\frac{2M}{r}-\frac{R r^2}{12},
\label{A-metric}
\end{equation}
where $M$ is a free parameter (called the mass parameter) and the value of $R$ is related to the cosmological constant by $R=4\Lambda $. If $M\neq 0$, the geometry is singular at $r=0$ and the position of the horizons, determined by the zeros of $A(r)$, are given by the positive real roots of a third degree polynomial. When $R \le 0$, there is only the event horizon with area radius $r_h$; for $0<R<4/(9M^2)$, besides the event horizon with radius $r_h$, there exists a cosmological horizon with radius $r_c$, which fulfills $r_h < r_c$. If $M=0$, the equation $A(r)=0$ is quadratic; no horizons exist when $R\le 0$ and only the cosmological horizon is present for any $R>0$, with $r_c=\sqrt{12/R}=\sqrt{3/\Lambda}$ in this case. 

For the inner region $\mathcal{M}_1$ of $\mathcal{M}$, we take a constant scalar curvature $R_1=4\Lambda_1$ and a vanishing $M_1=0$, while for the outer one $\mathcal{M}_2$ we choose a constant scalar curvature $R_2=4\Lambda_2$ and a non-zero $M_2: =M \neq 0$. The matching hypersurface area radius $a$ is taken so that no event horizons are present in both regions of $\mathcal{M}$, and when $R_1>0$ the region including the cosmological horizon of the inner region is also removed. As a consequence, the following constraints should be satisfied
\begin{equation}
1-\frac{a^2 R_1}{12}>0
\label{constra1}
\end{equation}
and
\begin{equation}
1-\frac{2 M}{a}-\frac{a^2 R_2}{12}>0.
\label{constra2}
\end{equation}
We can interpret the whole manifold $\mathcal{M}$ as having a singular hypersurface surrounding a static de Sitter region with a cosmological constant $\Lambda_1=R_1/4$, without matter and no horizons. Outside the shell the geometry represents a Kottler (or Schwarzschild-de Sitter) spacetime with a different cosmological constant $\Lambda_2=R_2/4$, no matter, and no event horizons. When $0<R_2<4/(9M^2)$, there exists a cosmological horizon in the outer region; in this case, the radius $a$ is smaller than the one corresponding to the cosmological horizon. Therefore, the presence of matter is only allowed at the matching hypersurface $\Sigma $, which then represents a kind of bubble. The construction should satisfy the condition given by Eq. (\ref{CondGen}), which can be written in the form
\begin{equation}
\left(4 -\frac{6 M}{a} -\frac{a^2 R_2}{2} \right) \sqrt{1-\frac{a^2 R_1}{12}}  -\left(4-\frac{a^2 R_1}{2}\right)\sqrt{1-\frac{2 M}{a}-\frac{a^2 R_2}{12}} = 0.
\label{ConGen2}
\end{equation}
From Eqs. (\ref{enden}) and (\ref{pres}), we obtain that
\begin{equation}
\sigma = \frac{1+2 R_1 \alpha }{\kappa} \frac{a R_1}{12} \left( \sqrt{1-\frac{a^2 R_1}{12}}\right)^{-1} +  \frac{1+2 R_2 \alpha }{\kappa} \left(\frac{ M}{a^2}-\frac{a R_2}{12}\right) \left(\sqrt{1-\frac{2 M}{a}-\frac{a^2 R_2}{12}}\right)^{-1}
\label{enden2}
\end{equation}
and
\begin{equation}
p =  \frac{1+2 R_1 \alpha }{\kappa a} \sqrt{1-\frac{a^2 R_1}{12}}-  \frac{1+2 R_2 \alpha }{\kappa a} \sqrt{1-\frac{2 M}{a}-\frac{a^2 R_2}{12}},
\label{pres2}
\end{equation}
and from Eq.  (\ref{T}) we find that
\begin{equation}
\mathcal{T} = \frac{ \alpha [R]}{\kappa a} \left(4-\frac{a^2 R_1}{2}\right) \left( \sqrt{1-\frac{a^2 R_1}{12}}\right)^{-1}
\label{T2a}
\end{equation}
or equivalently
\begin{equation}
\mathcal{T} = \frac{ \alpha [R] }{\kappa a } \left(4 -\frac{6 M}{a} -\frac{a^2 R_2}{2} \right) \left(\sqrt{1-\frac{2 M}{a}-\frac{a^2 R_2}{12}}\right)^{-1}.
\label{T2b}
\end{equation}
As shown before, the external energy flux vector is zero ($\mathcal{T}_\mu =0$), the dipole distribution strength is given by Eq. (\ref{dipo}), and it is non zero if $R_1 \neq R_2$. A particularly interesting case, which is actually the main goal of this paper, corresponds to a pure double layer, to be presented in detail next.

\section{Pure double layers}\label{pure}

As we have already discussed, a pure double layer must satisfy conditions (\ref{pure1}) and (\ref{pure2}) while keeping $[R]\neq 0$. From the previous construction, all the conditions in Eq. (\ref{pure1}) are already met, except for the first one which amounts to taking $\Sigma$ as a minimal timelike hypersurface on both sides, or equivalently, to setting $\mathcal{T}=0$. Thus, by using Eqs. (\ref{T2a}) and (\ref{T2b}) we impose that
\begin{equation}
R_1=\frac{8}{a^2}
\label{R1}
\end{equation}
and
\begin{equation}
R_2=\frac{8}{a^2} - \frac{12 M}{a^3}. 
\label{R2}
\end{equation}
For these values, the condition (\ref{constra1}) is always satisfied, but the inequality (\ref{constra2}) requires that 
\begin{equation}
a>3M \label{a3M}
\end{equation}
that we assume henceforth. This condition would not restrict $a$ if the parameter $M$ were negative. However, we are going to prove later (see Sec. \ref{M>0}) that the total quasi-local energy of the model as seen from the exterior $\mathcal{M}_2$ is proportional to $M$ with a positive constant of proportionality. Therefore, in order to keep the energy positive we assume that $M>0$ from now on. This sign of $M$ is actually related to the balance of the cosmological constants, as follows trivially from  Eqs. (\ref{R1}) and (\ref{R2}):
$$
\frac{[R]}{4} = [\Lambda ] = -\frac{3M}{a^3} 
$$
so that a positive $M$ implies a deficit in the outer cosmological constant with respect to the inner one.

It is obvious from Eq. (\ref{R1}) that $R_1 >0$, and then the combination of Eq. (\ref{a3M}) with Eqs. (\ref{R1}) and (\ref{R2}) provides
\begin{equation}
 0 < \frac{R_1}{2} < R_2 < R_1 \, . \label{ineq1}
\end{equation}
Furthermore, expression (\ref{R2}) implies that $R_2 (a=3M)=4/(9M^2)$ and also that $R_2$ is a decreasing function of $a$ on the allowed range $a\in(3M,\infty)$; hence we deduce that
\begin{equation}
R_2 < \frac{4}{9M^2} .\label{R2limit}
\end{equation}
This implies that the cosmological horizon is always present in the outer region.

Concerning conditions (\ref{pure2}), the last in them is simply $S_{\mu\nu}=0$ or, in other words, that $\sigma $ and $p$ are both zero. By replacing Eqs. (\ref{R1}) and (\ref{R2}) in Eq. (\ref{pres2}) and taking $p=0$, we obtain an expression that relates $a$ and $M$ with the quadratic theory coefficient $\alpha$:
\begin{equation}
-8 \alpha \left( 3 - \frac{3M}{a} + \sqrt{1-\frac{3M}{a}}\right) = a^2,
\label{alp1}
\end{equation}
so that we deduce that $\alpha$ must be negative. From Eq. (\ref{alp1}) we can also see by using the inequality (\ref{a3M}) that 
$$
 -16 \alpha < a^2 <-32 \alpha  \, .
$$
By introducing here Eq. (\ref{R1}) this inequality becomes
\begin{equation}
-\frac{1}{2} < \alpha R_1 < -\frac{1}{4}  \, . \label{des1}
\end{equation}

\begin{figure}[t!]
\centering
\includegraphics[width=0.5\textwidth]{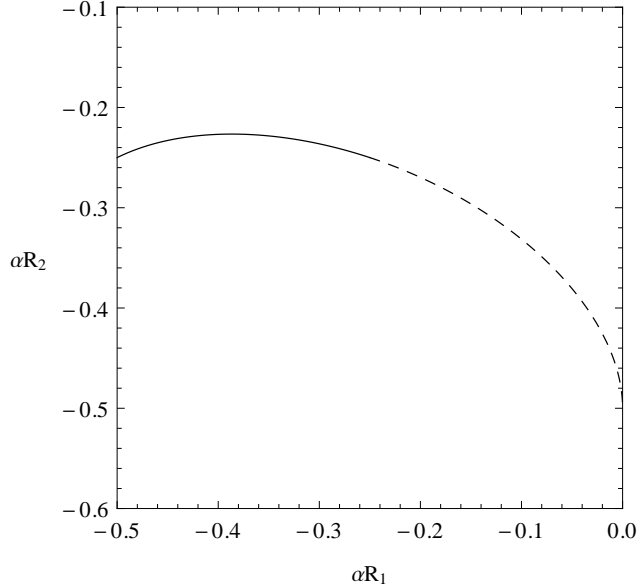}
\caption{\footnotesize{Pure double layer bubbles in quadratic $F(R)$. Plot of $\alpha R_2$ as a function of $\alpha R_1$, with $R_1$ and $R_2$ the scalar curvatures, and $\alpha$ the quadratic coefficient of the theory. The values of $R_1$ and $R_2$ are positive while $\alpha$ is negative (see text). The allowed values for $R_1$ and $R_2$ compatible with $M>0$ and a positive quasi-local energy are shown by the solid line. The continuation shown with the dashed line would be the feasible if $M<0$ were permitted, but this leads to negative energies as proven in subsection \ref{M>0}.}}
\label{fig1}
\end{figure}

\begin{figure}[t!]
\centering
\includegraphics[width=0.45\textwidth]{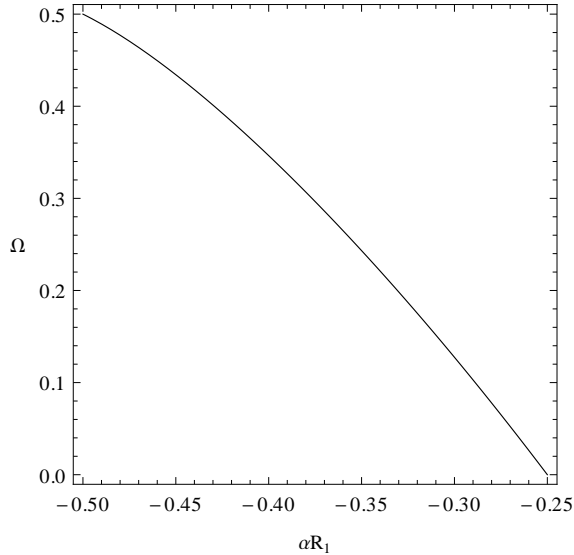}
\caption{\footnotesize{Pure double layer bubbles in quadratic $F(R)$. The dipole distribution strength has components $\kappa \mathcal{P}_{\hat{\imath} \hat{\jmath}}=\Omega h_{\hat{\imath} \hat{\jmath}}$, with $\Omega = 2\alpha [R]$ shown as a function of $\alpha R_1$.}}
\label{fig2}
\end{figure}

At this point, it is clear that Eqs. (\ref{ConGen2}) and (\ref{enden2}) are automatically satisfied ---and the same holds for the first in (\ref{pure2}). Finally, from Eq. (\ref{dipo}), the components of the dipole strength are given by $ \kappa \mathcal{P}_{\hat{\imath}\hat{\jmath}} = \Omega h_{\hat{\imath}\hat{\jmath}}$, with
\begin{equation}
\Omega = 2[R]\alpha = -\frac{24M}{a^3}\alpha 
\label{ome0}
\end{equation}
which has the same sign as $M$ and can also be expressed in terms of only $M$ and $a$ as
\begin{equation}
\Omega = \frac{3M} {3a-3M +\sqrt{a(a-3M})}.
\label{ome1}
\end{equation}
It is more appropriate for the physical interpretation to express the results in terms of $\alpha$, $R_1$, and $R_2$. We can invert Eqs. (\ref{R1}) and (\ref{R2}) to find the bubble area radius $a$ and the parameter $M$:
\begin{equation}
a=\frac{2\sqrt{2}}{\sqrt{R_1}}, 
\label{apdl}
\end{equation}
and
\begin{equation}
M = \frac{4\sqrt{2}(R_1 - R_2)}{3R_1\sqrt{R_1}}. 
\label{Mpdl}
\end{equation}
By replacing these values of $a$ and $M$ in Eq. (\ref{alp1}), we find that  $R_1$ and $R_2$ are related with $\alpha$ by the equation
\begin{equation}
-\frac{1}{\alpha} =  R_1 + 2 R_2 + \sqrt{R_1(2R_2-R_1)} .
\label{alp2}
\end{equation}
Hence, if the parameter $\alpha$ of the quadratic theory  is given and is negative, we can choose one of the cosmological constants within the allowed limits, and then the other is immediately fixed by relation (\ref{alp2}). Thus, once $\alpha <0$ is fixed, one can choose any value of the scalar curvature $R_1$ within the allowed range (\ref{des1}) and obtain the associated value of the scalar curvature $R_2$ from Eq. (\ref{alp2}), which gives
\begin{equation}
R_2 = \frac{-2 - \alpha R_1 + \sqrt{-\alpha R_1 (4 + 7 \alpha R_1) }}{4 \alpha}.
\label{R2R1}
\end{equation}
The corresponding values of the bubble radius $a$ and the mass $M$ are calculated from Eqs. (\ref{apdl}) and (\ref{Mpdl}), respectively. The dipole strength $\Omega$ is then found by combining Eqs. (\ref{ome1}) and (\ref{R2R1}):
\begin{equation}
\Omega = \frac{-2 - 5 \alpha R_1 + \sqrt{-\alpha R_1 (4 + 7 \alpha R_1)}}{2}.
\label{ome2}
\end{equation}
A similar but more complicated analysis can be done in terms of $R_2$ instead of $R_1$. The results are graphically shown in Figs. \ref{fig1} and \ref{fig2}. For the conformal diagrams of the solution, see Sec. \ref{diagrams}.

\subsection{Quasilocal energy and the sign of $M$}\label{M>0}

In general relativity we know that $M$ represents the total mass of the body creating the spacetime $\mathcal{M}_2$ and can be given quasi-local descriptions as the Misner-Sharp mass or the Hawking quasi-local energy among many others \cite{Sza}. The situation is different in principle in $F(R)$ theories, and one can prove \cite{ccho} that the correct generalization for a quadratic theory with Lagrangian (\ref{quadF}) is given by the quasi-local energy
\begin{equation}
E = (1+2\alpha R) M \, .\label{E}
\end{equation}
The question arises whether one can have a negative mass parameter $M<0$ while keeping a positive energy $E$. This might be possible in general, as long as $2\alpha R <-1$, but it is not feasible in our pure double layers due to the strict relation between $R_1$ and $R_2$ at both sides. To prove it, observe that $E_1=0$, and thus we only need to consider $E_2 =(1+2\alpha R_2)M$. The allowed range of $\alpha R_1$, given by inequality (\ref{des1}) for $M>0$, extends to $-1/2<\alpha R_1 <0$ if we also let $M<0$. In this case, formula (\ref{R2R1}) ---which is valid independently of the sign of $M$--- implies that  $\alpha R_2$ is smaller than a certain (negative) maximum and always larger than $-1/2$, as shown in Fig. \ref{fig1}. Therefore, $1+2\alpha R_2$ is positive for the pure double layers, and the sign of $E_2$ is necessarily the same as the sign of $M$. This restricts the analysis to the case with $M>0$.

We can actually provide a possible physical interpretation of the dipolar strength $\Omega$ in terms of the quasi-local energy $E$. The difference between the energies at both sides of the double layer is 
$$E_2- E_1 = E_2 = (1+2\alpha R_2) M= M(1+16\alpha/a^2 -24\alpha M /a^3) =\tilde{E} +\Omega M
$$
where $\tilde{E}= M(1+2\alpha R_1)$ is the energy that the double layer would have at $\mathcal{M}_1$, that is, if the cosmological constants did not jump. A kind of perfect balance seems to exist between the cosmological constants and the presence of the pure double layer with its dipole strength.

\subsection{Extension across the cosmological horizons: Conformal diagrams}\label{diagrams}

\begin{figure}[t!]
\centering
\includegraphics[width=0.95\textwidth]{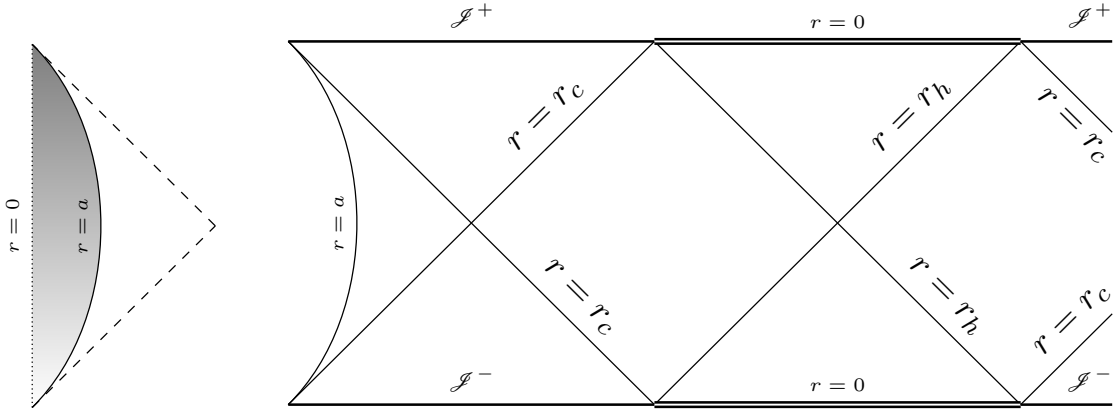}
\caption{\footnotesize{Conformal diagrams of the two regions $\mathcal{M}_1$ and $\mathcal{M}_2$ composing the total spacetime. As usual null radial lines are at 45$^o$. On the left picture we show the portion of static de Sitter spacetime describing the inner part of the bubble. The origin of coordinates $r=0$ is represented by the dotted line, while the pure double layer is placed at $r=a$. Only the shaded zone describes the inner region $\mathcal{M}_1$, and this has to be joined to the exterior $\mathcal{M}_2$ depicted on the right picture, which represents the conformal diagram of the portion of Schwarzschild-de Sitter spacetime describing the outer part $\mathcal{M}_2$ of the bubble. This spacetime ``starts'' at the timelike hypersurface $r=a$ ---representing the pure double layer--- within a static region. Then, the metric can be extended, to the future and to the past, across the cosmological horizon labeled $r=r_c$, leading to past and future null infinities $\scri^\pm$, and also to a new event horizon $r=r_h$ that encloses the singularities shown by double lines and marked as $r=0$. The metric can then be extended towards ``the right'' indefinitely, by alternating an infinite number of cosmological and event horizons. As we see, such a bubble only removes some of the many singularities of the Kottler spacetime, and therefore it might be better to cut the outer part by placing a second, symmetric, bubble, before the first event horizon appears. This is represented in Fig. \ref{fig4}.}}
\label{fig3}
\end{figure}

\begin{figure}[t!]
\centering
\includegraphics[width=0.5\textwidth]{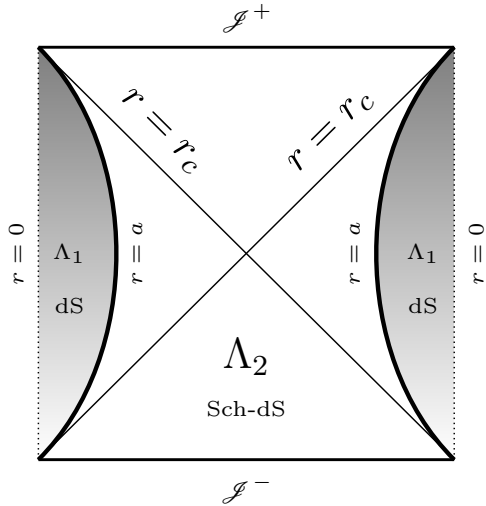}
\caption{\footnotesize{The conformal diagram of the total bubble, avoiding any curvature singularities but allowing for distributional curvature tensors. This metric is obtained by joining the two portions shown in Fig. \ref{fig3} across the hypersurfaces labeled $r=a$ as explained in the main text, and by then doing a similar procedure on the symmetric ``right" part of the cosmological horizon. The two shaded parts are equal copies of the portion $r<a$ of static de Sitter spacetime with $\Lambda_1 >0$, and the non-shaded part is a portion of Kottler spacetime with $\Lambda_2\in (\Lambda_1/2, \Lambda_1)$, mass parameter $M=(\Lambda_1 -\Lambda_2)a^3/3$ and without any event horizons ---so that the curvature singularities have been removed. There are pure double layers on the hypersurfaces $r=a$ drawn with thick lines, having total strength $\Omega =-8\alpha (\Lambda_1 -\Lambda_2)=-24\alpha M/a^3$ which is positive ---as $\alpha$ is required to be negative. In the limit $\Lambda_2 \rightarrow \Lambda_1$ the pure double layer disappears, $M$ vanishes and one ends up with the entire de Sitter spacetime.}}
\label{fig4}
\end{figure}

From our previous discussion we know that there are no horizons in the inner region $\mathcal{M}_1$ but, given that Eq. (\ref{R2limit}) always holds, there appears a cosmological horizon on the outer region of the bubble $\mathcal{M}_2$. This horizon is placed at $r=r_c$ with $r_c$ defined by
$$
1-\frac{2M}{r_c} -\frac{R_2}{12} r^2_c =0 .
$$
Our coordinates do not describe the region beyond the cosmological horizon, with $r>r_c$, but a standard extension can be obtained by simply defining Eddington-Finkelstein-type advanced/retarded coordinates
$$
v_\pm = t\pm \int\frac{dr_2}{A_2^2(r_2)} 
$$
so that the metric on $\mathcal{M}_2$ becomes 
$$
ds^2 =-A_2(r_2) dv_\pm^2 \pm dr_2  dv_\pm +r_2^2 \left(d\theta^2 +\sin^2\theta d\varphi^2 \right).
$$
By gluing different patches of this type one can obtain a maximal extension in the standard way. However, if we performed such an extension we would end up with a typical spacelike singularity of Schwarzschild type with $r=0$ beyond the cosmological horizon, and there would appear an infinite number of event and cosmological horizons as in the usual Kottler metric with positive cosmological constant. This is graphically explained in the conformal diagrams presented in Fig. \ref{fig3}. 

Given that we wanted to remove the singularities and event horizons from the inner region, it does not seem acceptable to have this kind of behavior elsewhere, and therefore we propose a different, more interesting and symmetrical possibility. This is shown in Fig. \ref{fig4}. The basic idea is to place another bubble when the area radius becomes $r=a$ again, with exactly the same junction conditions, so that the region with $r<a$ is again replaced by a second, ``mirror" symmetric, bubble with the same properties. The total spacetime is thus free of curvature singularities but including pure double layers at $r=a$, the junction between the different cosmological constants. The total spacetime can be thought of as having two, antipodean, static de Sitter bubbles with a Kottler portion ---a portion without singularities--- in between. Alternatively, one could try to make identifications between the two mirror bubbles to produce a unique one.

\section{Discussion}\label{discu}

In this paper, we have constructed spherical bubbles of constant area radius $a$ in quadratic $F(R)$ gravity with Lagrangian (\ref{quadF}) by using the corresponding junction formalism. We have adopted for the inner region a vacuum solution with constant scalar curvature $R_1=4\Lambda_1$ and no mass, and for the outer one another vacuum solution with constant scalar curvature $R_2 =4\Lambda_2$ and non-zero mass parameter $M$, leading to non-zero quasi-local energy $E_2=(1+2\alpha R_2)M $. The matching conditions result in the presence of a thin shell, where the energy-momentum tensor is singular (it has a distributional part), characterized by a hypersurface energy density and a pressure. When $R_1 \neq R_2$, there is also an external scalar tension/pressure and a double layer tensor distribution, with a strength proportional to the jump of $R$ across the shell and to the quadratic coefficient of the theory $\alpha $. A third additional contribution that appears when $R_1 \neq R_2$, corresponding to the external energy flux vector, is zero in our construction because both scalar curvatures are chosen to be constant. 

The possibility of the existence of pure double layers is a remarkable property that distinguishes quadratic $F(R)$ gravity from other (non quadratic) $F(R)$ theories and from general relativity. We have found that our bubble construction constitutes an explicit example where this interesting feature is present, whenever the quadratic coefficient $\alpha$ is negative. 

When $\alpha <0$, $-1/2 < \alpha R_1 < -1/4$, and $0 < R_1/2 < R_2 < R_1$, for a given value of $\alpha$, there is always a combination of the values of $R_1$ and $R_2$ for which the energy density and the pressure at the shell are both zero (and also the scalar external tension/pressure), but the dipole distribution strength $\kappa \mathcal{P}_{\hat{\imath}\hat{\jmath}} = \Omega h_{\hat{\imath}\hat{\jmath}}$ is not, resulting in a pure double layer. Once $\alpha<0$ is chosen, so that the gravity theory is fixed, these bubbles threaded by pure double layers have an area radius and a mass parameter determined by the value of $R_1$ (or $R_2$). The double layer is pure only when there is a fine tuning between the parameters; if not, the thin shell gradually reappears. This feature can be visualized in Fig. \ref{fig1}: moving along the curve maintains the double layer pure, but outside it the thin shell is present, with non-vanishing energy density and pressure. This is reminiscent of the analogous behavior in classical electrodynamics \cite{J}, where {\em pure} dipole layers arise when the charge surface densities at both sides of the layer are of opposite sign but equal in absolute value; otherwise, a surface charge density also arises. Therefore, this is not necessarily an indication of instability of our pure gravitational double layers.

In our model, the construction of these simple bubbles is possible in any quadratic $F(R)$ gravity, but pure double layers may appear only if the theory has a negative $\alpha$. In this case, the energy-momentum tensor of the spacetime takes the distributional form
$$
T_{\mu\nu} =8\frac{\alpha}{\kappa} (\Lambda_2-\Lambda_1)  \nabla_\rho \left(h_{\mu\nu} n^\rho \delta^\Sigma \right)
$$
where $\delta^\Sigma$ is the Dirac delta supported on the hypersurfaces $\Sigma : \{ r=a\}$ and $h_{\mu\nu}$ is the first fundamental form on $\Sigma$. Hence, $T_{\mu\nu}$ is supported only on $\Sigma$; that is, it vanishes everywhere except at $r=a$, describing there a dipole-like  ---in the sense of a Dirac-delta derivative--- source. In these pure double layer spacetimes, the exterior region has a deficit in the cosmological constant with respect to the interior one, due to the presence of the double layer at $\Sigma$. This deficit is related with the mass parameter by $M/a^3=(\Lambda_1-\Lambda_2)/3$.  It is also remarkable that the construction of this pure double layer requires $\Sigma$ to be a minimal hypersurface, in direct analogy with the minimal surfaces describing classical soap bubbles. 
 
On the other hand, it is well known \cite{revfr1} that any $F(R)$ gravity theory is equivalent to a certain scalar-tensor theory. In particular, quadratic $F(R)$ is equivalent to the Brans-Dicke theory with parameter $\omega = 0$; in this case, the scalar field $\phi$ is related with $R$ by $\phi = 2\alpha R -1$, with a potential $V(\phi) =2\Lambda +(\phi^2-2\phi-3)/(4\alpha)$ \cite{revfr1}. From this point of view, the double layer corresponds to an abrupt discontinuity in the scalar field $\phi$. In our bubble construction, the scalar field would have different constant values $\phi _1$ and $\phi _2$ at each side of the matching surface, which can be interpreted as some kind of effective gravitational constants. In the special case of pure double layers, it is interesting to note that the discontinuity in the scalar field, given by $[\phi]=2\alpha [R]$, is precisely the value $\Omega$ corresponding to the dipole strength of the double layer.

It is usually claimed that a quantum theory of gravity requires the presence of the quadratic terms in the Lagrangian; in this case, double layers like those discussed in this work may arise as an idealized representation of situations where the cosmological constant has an abrupt jump. However, the shell matching in a generic quadratic gravity (not $F(R)$) requires the continuity of the second fundamental form ---which is the proper matching in general gelativity (GR)--- and in this sense, as pointed out in previous articles \cite{js1,js2,js3,rsv}, hypersurfaces in GR with a proper matching would actually become double layers in any regime where quadratic terms start to be non negligible. For example, when approaching a quantum regime in which quadratic terms become significant, any classical GR solution that can be approximately described by a matching procedure will become an approximate solution which contains, on the matching hypersurface, a gravitational double layer. On the other hand, in quadratic $F(R)$ the matching allowing for shells only fixes the continuity of the trace of the second fundamental form, and thus GR thin shells of matter or some classes of less conventional theoretical objects, such as GR braneworlds, domain walls, gravastars, and wormholes may become double layers if the quadratic term cannot be ignored. In particular, as we have shown for the case of the simple bubbles studied here, pure double layers may appear in some of these models if the relevant parameters are properly fine tuned. The simple model we have presented herein can also be considered as a valid spacetime in GR, where it represents a thin shell with a traceless energy-momentum tensor. It would be very interesting to find pure double layers in non-$F(R)$ quadratic gravity, where the models would be properly matched solutions in GR but pure double layers when the quadratic terms start to dominate. This is an open problem that deserves further consideration.

\section*{Acknowledgments}

E.F.E. and G.F.A. are supported by CONICET and Universidad de Buenos Aires. J.M.M.S. is supported under grants FIS2014-57956-P (Spanish MINECO-fondos FEDER), IT956-16 (Basque Government), and EU COST action CA15117 ``CANTATA''.

\end{document}